\documentclass[aps,prl,twocolumn,superscriptaddress,notitlepage]{revtex4-2}
\usepackage{graphicx}
\usepackage{dcolumn}
\usepackage{bm}
\usepackage{amsmath,amssymb,amsthm,graphicx, color}
\usepackage{times}
\usepackage{dsfont}
\usepackage{braket}
\usepackage{multirow}
\usepackage[colorlinks,allcolors=blue]{hyperref}
\usepackage[utf8]{inputenc}
\usepackage{dcolumn}
\usepackage{bm}
\usepackage{times}
\usepackage{float}

\begin{document}

\title{Magnetization-free quantum anomalous Hall effect in altermagnetic weak topological insulators}

\author{Yanmiao Han}
\affiliation{School of Physics, Beihang University, Beijing 102206, China}

\author{Yu-Hao Wan}
\email[Contact author:] {wanyh@stu.pku.edu.cn}
\affiliation{International Center for Quantum Materials and School of Physics, Peking University, Beijing 100871, China}
\author{Peng-Yi Liu}
\affiliation{International Center for Quantum Materials and School of Physics, Peking University, Beijing 100871, China}

\author{Rundong Zhao}
\email[Contact author:] {rdzhao@buaa.edu.cn}
\affiliation{School of Physics, Beihang University, Beijing 102206, China}

\author{Qing-Feng Sun}
\email[Contact author:] {sunqf@pku.edu.cn}
\affiliation{International Center for Quantum Materials and School of Physics, Peking University, Beijing 100871, China}
\affiliation{Hefei National Laboratory, Hefei 230088, China}

\begin{abstract}
The quantum anomalous Hall (QAH) effect is usually realized through ferromagnetism or magnetic doping, which limits material choices and often results in low mobility, disorder sensitivity, and low operating temperatures.
Here, we propose a mechanism to achieve the QAH phase without net magnetization, using altermagnetic exchange on a weak topological insulator (WTI).
The altermagnetic order selectively gaps the Dirac cone at the $Z$ point,
while hybridization between the top and bottom surfaces of the WTI thin film
opens a trivial gap at $\Gamma$ point, generating a Chern number $\mathcal{C}=1$ phase with a single chiral edge channel.
We verify the topological transition through both a minimal thin-film model and a slab geometry, and compute the associated Hall response.
As altermagnets and layered WTIs already exist experimentally, this mechanism offers a promising and experimentally accessible route to realizing QAH phases without magnetic doping or global magnetization. The results indicate that altermagnetic surface engineering may provide a clean and controllable pathway for realizing dissipationless edge transport in layered quantum materials.
\end{abstract}

\maketitle
\textit{Introduction. —}
The quantum anomalous Hall (QAH) effect yields a quantized Hall conductance in the absence of an external magnetic field, driven purely by intrinsic band topology and magnetic order, and has attracted enormous attention over the past decade\cite{RevModPhys.95.011002,yu2010quantized,chen2020tunable,PhysRevLett.101.146802,RevModPhys.82.3045,8vs2-jvc4,PhysRevB.87.081105,PhysRevB.85.115439,addr1,PhysRevB.95.045113,addr2,PhysRevLett.61.2015,PhysRevLett.125.126801,mogi2015magnetic,PhysRevMaterials.4.111201,PhysRevLett.113.137201,PhysRevB.95.121410}. Following the prediction that magnetic doping of three-dimensional (3D) topological insulators can produce a Chern-insulating state\cite{yu2010quantized},
the QAH effect was first observed in Cr- and V- doped $(\mathrm{Bi}\mathrm{Sb})_{2}\mathrm{Te}_{3}$ thin films\cite{Chang2013QAH}.
More recently, QAH phases have been realized in additional platforms, including exfoliated flakes of the intrinsic magnetic topological insulator (TI) $\mathrm{MnBi}_{2}\mathrm{Te}_{4}$\cite{doi:10.1126/science.aax8156} and moir\'e superlattices \cite{doi:10.1126/science.aay5533}, such as twisted bilayer graphene aligned to hexagonal boron nitride\cite{doi:10.1126/science.aaw3780}, rhombohedral (ABC) trilayer graphene on $h$-BN\cite{PhysRevB.111.L081111}, and transition-metal dichalcogenide heterobilayers like AB-stacked $\mathrm{MoTe}_{2}/\mathrm{WSe}_{2}$\cite{li_quantum_2021}.
Nevertheless, achieving the QAH effect still largely relies on ferromagnetism, which is typically realized via magnetic doping or spontaneous magnetization. Ferromagnetism can induce stray fields, causing interference between devices and significantly affecting their integration density. Therefore, realizing the QAH effect in non-ferromagnetic materials has important application potential\cite{PhysRevLett.134.116603}.

A promising direction is altermagnetism, a recently identified class of collinear magnetism with zero net magnetization yet large, momentum-dependent spin splitting enforced by crystal symmetries\cite{PhysRevX.12.040501,PhysRevX.12.031042,PhysRevB.110.144412,PhysRevLett.131.076003,PhysRevB.110.024503}. Despite vanishing macroscopic magnetization, altermagnetic order can generate spontaneous Hall responses\cite{vsmejkal2020crystal, PhysRevLett.130.036702,PhysRevLett.134.136301,PhysRevB.111.045409,vsmejkal2022anomalous,feng_anomalous_2022} and can host nontrivial topological band structures\cite{PhysRevB.109.024404,PhysRevB.110.024425,PhysRevB.108.184505,PhysRevB.110.094425,PhysRevB.109.224430,PhysRevLett.134.096703,PhysRevB.110.L220402, PhysRevB.111.045106, s57q-q7gt}.
These features indicate that altermagnetism could potentially realize a QAH phase without global ferromagnetism and provide a route to engineering Chern-insulating states.

In 3D topological band theory, TIs are classified as strong or weak, characterized by an odd or even number of surface Dirac cones, respectively\cite{RevModPhys.83.1057,PhysRevLett.98.106803,PhysRevB.76.045302,PhysRevB.75.121306}. Magnetic phenomena in strong TIs have been extensively explored and yield numerous magnetic topological phases\cite{jain2024quantized,liu2020robust,PhysRevB.108.L241407,mogi2022experimental}. In contrast, weak TIs (WTIs) remain comparatively under-explored under magnetic order, and how such order reshapes their surface states and enables Hall-type phases is still largely open.

In this Letter, we propose a magnetization-free mechanism for the QAH effect: a thin film of a 3D WTI becomes a Chern insulator when altermagnetic exchange is applied on its two surfaces, even though the system carries no net magnetization. This establishes that QAH physics need not rely on ferromagnetic strong TIs and highlights altermagnetic surface engineering as a viable route toward dissipationless edge transport.

\begin{figure}[htbp]
    \includegraphics[width=1\linewidth]{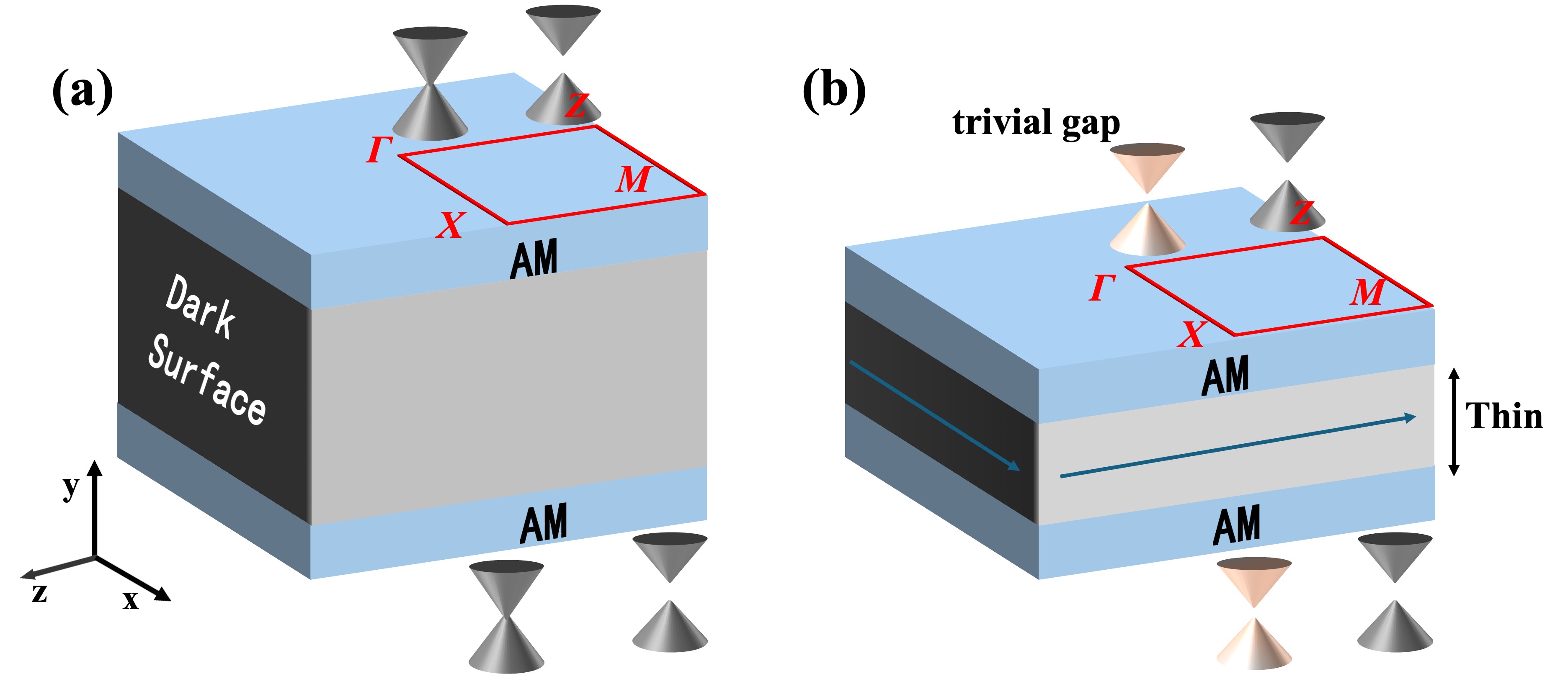}
        \caption{Schematic illustration of an altermagnetic (AM) WTI. (a) A 3D WTI with $\mathbb{Z}_4$ index $(0,001)$: the $(001)$ side is a dark surface without gapless Dirac cones, while the top and bottom AM surfaces host two Dirac cones at $\Gamma$ and $Z$, with surface altermagnetism gapping out one cone at point $Z$ and leaving a single gapless Dirac cone at point $\Gamma$. (b) A 3D WTI thin-film: hybridization between the top and bottom surfaces opens an additional trivial gap at $\Gamma$, while the gap generated by altermagnetic remains nontrivial and gives rise to chiral edge modes propagating along the side surface.
    }
        \label{fig1}
\end{figure}

As illustrated in Fig.\hyperref[fig1]{1(a)}, we start from a 3D WTI characterized by a $\mathbb{Z}_4$ index $(0,001)$, which represents a broad class of layered WTIs realized in materials such as Bi$_4$Br$_4$ and Bi$_2$TeI, as well as other bismuth halides\cite{eschbach2017bi1te1,majhi2017emergence,rusinov2016mirror}. When altermagnetism is imposed on the top and bottom surfaces, one of the two gapless Dirac cones on each surface is gapped out at the $Z$ point, leaving a single gapless Dirac cone at the $\Gamma$ point on the surface. As the film becomes thinner (Fig.\hyperref[fig1]{1(b)}), hybridization between the top and bottom surfaces opens an additional gap at the $\Gamma$ point. It is important to emphasize that this hybridization-induced gap at $\Gamma$ is topologically trivial, whereas the altermagnetism generates a nontrivial gap that supports a chiral edge state. We describe this system using both an effective surface model and a realistic slab model of a WTI with surface altermagnetism, and use these models to elucidate the emergence of the QAH effect.

\textit{Thin Film Model of altermagnetic Weak Topological Insulator. —}\label{sec2}
As a start for constructing the thin-film model, we consider the effective surface lattice Hamiltonian on the $(010)$ side surface of a 3D WTI characterized by the $\mathbb{Z}_4$ index $(0,001)$\cite{PhysRevB.111.045407}:
\begin{equation}
\begin{aligned}
     H_{surf} &= H_{Dirac}+H_{W}^{aniso}
       \\&= A(\sin k_z \sigma_x + \sin k_x \sigma_y) +2B_x(1-\cos k_x) \sigma_z  ,
\end{aligned}
\end{equation}
where $\sigma_{x/y/z}$ are the Pauli matrices in the spin space.
The lattice-regularized Hamiltonian $H_{\text{Dirac}} = A\bigl(\sin k_z\sigma_x + \sin k_x\sigma_y\bigr)$ describes a 2D Dirac fermion on the (010) surface of a square lattice (with lattice constant = 1), where $A$ is the Fermi velocity and throughout this work we set $A = 1$ for convenience.
$H_W^{\text{aniso}} = 2B_x (1-\cos k_x) \sigma_z$ is an anisotropic Wilson mass that ensures that the $(010)$ surface hosts only two Dirac cones located at the $\Gamma$ and Z points. The anisotropic Wilson term $H_W^\text{aniso}$ vanishes as $k_x$ approaches zero in a quadratic manner, so that the Dirac cones at $\Gamma$ and $Z$ remain gapless, whereas the Dirac cones at the $X$ and $M$ points acquire a finite mass gap.

Based on the above surface Hamiltonian $H_\text{surf}$, we construct the thin-film model. For a sufficiently thin film, the low-energy physics model is described by the following 4-band Hamiltonian in the basis $(|t\uparrow\rangle, |t\downarrow\rangle, |b\uparrow\rangle, |b\downarrow\rangle)$\cite{s6pj-495v, wang2015quantum, PhysRevB.81.115407}, where $t$ and $b$ refer to the top and bottom surfaces, and $\uparrow/\downarrow$ denote spin direction:
\begin{equation}
H_0 =
\begin{bmatrix}
 0       & i A k_{-} & m_k^* & 0 \\
 - i A k_{+} & 0     & 0      & m_k^* \\
 m_k       & 0             & 0 & - i A k_{-} \\
 0            & m_k        & i A k_{+} & 0
\end{bmatrix}.
\end{equation}
Here, $k_{\pm} = \sin k_x \pm i \sin k_z$, and $m_k = m_{0}-2B_{x}^{'}(1-\cos k_{x})-2B_{z}^{'}(1-\cos k_{z})$ denotes the hybridization-induced mass term. 

Since we are dealing with the (010) surface of a 3D WTI, we need to add an anisotropic Wilson term $H_W^{aniso}= 2B_x (1-\cos k_x) \tau_{0}\otimes \sigma_{z}$ to regularize the lattice, so that the surface hosts Dirac cones only at the $\Gamma$ and $Z$ points. Here, $\tau_{0}$ is the unit matrix in the top/bottom space.
To explore the effect of altermagnetic order, we introduce an altermagnetic term,
$H_{\mathrm{altermag}}=J(\cos k_{x}-\cos k_{z})\tau_{0}\otimes \sigma_{z}$,
where $J$ characterizes the strength of the altermagnetic order. Unlike a uniform Zeeman field, $H_{\mathrm{altermag}}$ is a momentum-dependent interfacial exchange field. Its $d_{x^2-z^2}$ form factor $\cos k_x-\cos k_z$ has zero Brillouin-zone average, producing no ferromagnetic component while breaking time-reversal symmetry, and corresponds to the $\mathcal{C}_{4}\mathcal{T}$-symmetric altermagnetic mass discussed for two-dimensional Dirac fermions~\cite{k6th-nyt9}. It is also valley selective: $H_{\mathrm{altermag}}(\Gamma)=0$, whereas $H_{\mathrm{altermag}}(Z)=2J$. After block diagonalization, the thin-film Hamiltonian becomes $h^{(\eta)}(\mathbf{k})=A(\sin k_z\sigma_x-\eta\sin k_x\sigma_y)+[m_{\mathbf{k}}+\eta M(\mathbf{k})]\sigma_z$, with $\eta=\pm$ and $M(\mathbf{k})=2B_x(1-\cos k_x)+J(\cos k_x-\cos k_z)$; see Supplemental Material~\cite{SupplementalMaterial}, Sec.~S1. This valley-selective mass generation is the microscopic origin of the magnetization-free QAH phase.

From a materials viewpoint, a possible route is to proximity couple layered bismuth-halide WTI or related topological-insulator films, such as Bi$_4$Br$_2$I$_2$, Bi$_4$Br$_4$, and Bi$_2$TeI~\cite{Noguchi2024Bi4Br2I2,Zhao2023Bi4Br4,Tang2014Bi2TeI}, to altermagnetic substrates or overlayers such as MnTe, CrSb, RuO$_2$, or Mn$_5$Si$_3$~\cite{Krempasky2024MnTeAltermagnetism,Yang2025CrSbAltermagneticSplitting,feng_anomalous_2022,PhysRevB.109.224430}. In such heterostructures, the altermagnetic proximity effect can imprint a momentum-dependent spin splitting onto the adjacent nonmagnetic WTI surface layers, thereby realizing the effective exchange term used in our model. This scenario is supported by recent first-principles work showing a proximity-induced scale of several tens of meV in an adjacent nonmagnetic layer~\cite{kqy8-myz1}.

\begin{figure}[htbp]
    \includegraphics[width=1\linewidth]{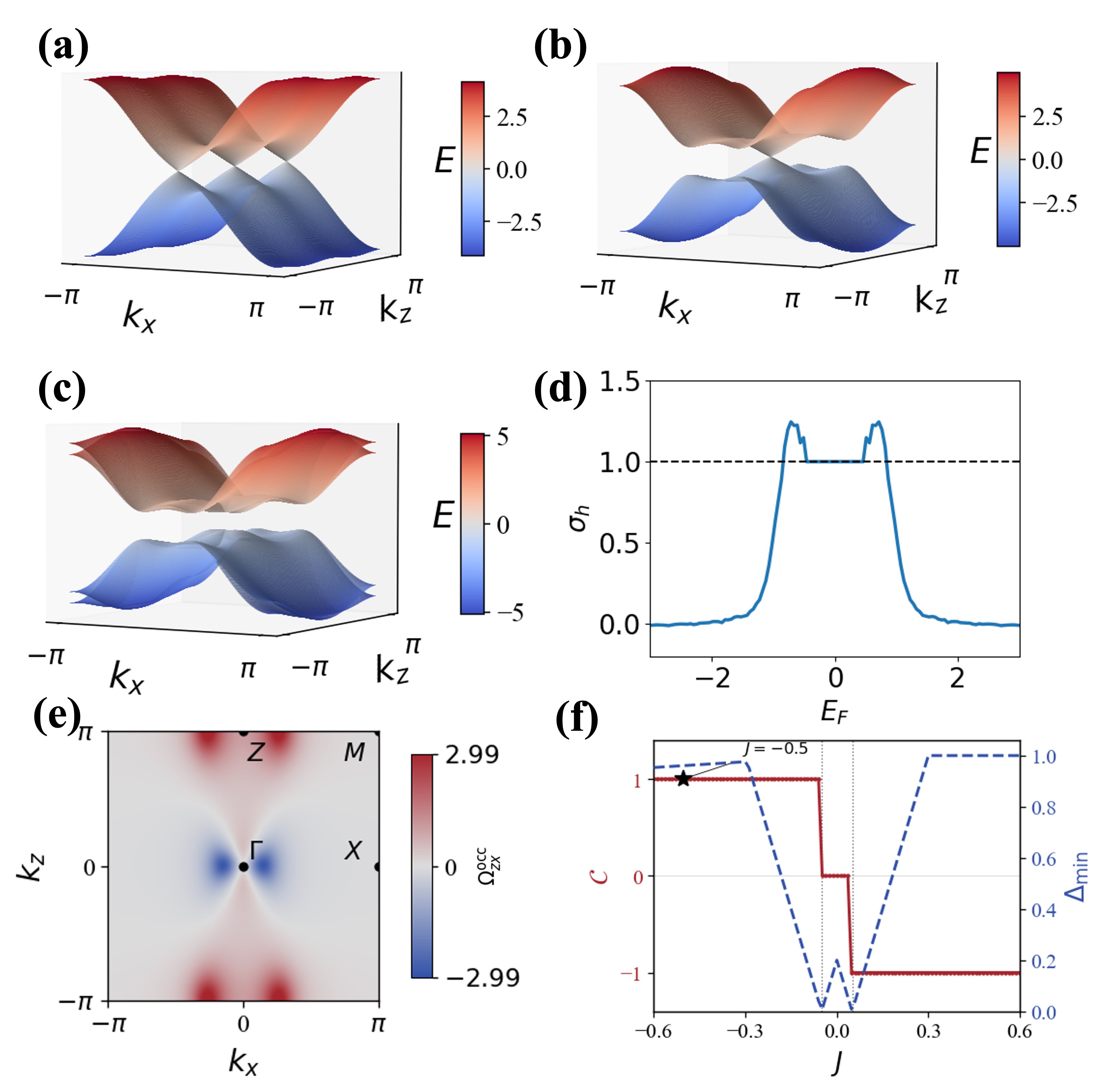}
        \caption{Band structures, Berry curvature, and Chern-number evolution of the
altermagnetic WTI thin film. 
	(a-c) Band structures without altermagnetism and without inter-surface hybridization ($J=0$, $m_k=0$) in (a),
	with altermagnetism $(J=-0.5)$ but without inter-surface hybridization in (b), and
	with both altermagnetism $(J=-0.5)$ and finite inter-surface hybridization $(m_0=0.5, B_x^{'}=B_z^{'}=0.1)$ in (c).
	(d) Hall conductivity for the parameters in (c), showing a quantized plateau when $E_F$ lies inside the gap (in units of $\frac{e^2}{h}$).
	(e) Occupied-band Berry curvature $\Omega_{zx}^{\rm occ}(k_x,k_z)$ for the parameters in (c). 
	(f) Minimum direct gap $\Delta_{\min}$ and Chern number $\mathcal{C}$ versus $J$.
	    }
        \label{fig2}
\end{figure}

Based on the thin-film model introduced above, we now calculate the band structure of a 3D WTI. 
Fig.\hyperref[fig2]{2(a)} shows the spectrum in the absence of altermagnetism ($J=0$) and without hybridization between the top and bottom surfaces ($m_k=0$). In this case, both blocks host gapless Dirac cones at the $\Gamma$ and $Z$ points, consistent with the characteristic even number of surface Dirac cones of a 3D WTI, and confirming that our thin-film model accurately captures the physics of the top and bottom surfaces.
In Fig.\hyperref[fig2]{2(b)} we turn on altermagnetic order ($J=-0.5$) while still neglecting hybridization. The altermagnetism induces a band inversion and opens a gap at $Z$, while the Dirac cones at $\Gamma$ remain gapless.
Figure~\hyperref[fig2]{2(c)} further includes finite inter-surface hybridization ($m_k \not=0$):
the $Z$-point gap remains robust, the degeneracy between the two surface-derived branches is partially lifted over the Brillouin zone, and the $\Gamma$ cones become gapped.
This $\Gamma$-point gap is topologically trivial and arises solely from inter-surface hybridization.

While the band structures in Figs.~\hyperref[fig2]{2(a)}--\hyperref[fig2]{2(c)} show the opening of Dirac-cone gaps, the band dispersion alone does not directly establish their topological character. Thus, using the same parameters as Fig.\hyperref[fig2]{2(c)},
we compute the Hall conductivity (see Supplemental Material~\cite{SupplementalMaterial}, Sec.~S2), shown in Fig.\hyperref[fig2]{2(d)}. When the Fermi energy lies inside the hybridization gap, the Hall conductivity develops a quantized plateau at $\sigma_h = 1$ (in units of $\frac{e^2}{h}$). As $E_F$ is further increased beyond the gap, this quantized Hall plateau disappears, and Hall conductivity gradually decreases toward zero.
Thus, the altermagnetic WTI thin film, with the Fermi level in the hybridization gap, exhibits a total Chern number $\mathcal{C} = 1$ (corresponding to a Hall conductivity $\sigma_{h} = \frac{e^2}{h}\mathcal{C}$), characteristic of the QAH effect. Figs.\hyperref[fig2]{2(e)} and \hyperref[fig2]{2(f)} further distinguish the two gaps in momentum and parameter space. Fig.\hyperref[fig2]{2(e)} shows the occupied-band Berry curvature \(\Omega_{zx}^{\rm occ}(k_x,k_z)\) after both gaps open. Although Berry-curvature weight appears near the hybridized \(\Gamma\) cone, the two occupied low-energy blocks have opposite Dirac chirality and cancel around \(\Gamma\), as derived in the Supplemental Material~\cite{SupplementalMaterial}. The relevant masses are \(m_{\eta}(\Gamma)=m_0\), independent of \(J\), and \(m_{\eta}(Z)=m_0-4B_z'+2\eta J\), which can change sign under altermagnetic exchange. Accordingly, Fig.\hyperref[fig2]{2(f)} shows a gap closing and reopening at the \(Z\)-point mass inversion, accompanied by a Chern-number jump. Thus, the QAH phase originates from the \(Z\)-point mass inversion, whereas the \(\Gamma\) gap is a topologically trivial hybridization gap; see Supplemental Material~\cite{SupplementalMaterial} (see also references~\cite{Fukui2005FHS} therein), Sec.~S3.

To further test the robustness of this mechanism and enable transport calculations, we next study a microscopic slab model with finite thickness.

\begin{figure}[htbp]
    \includegraphics[width=1\linewidth]{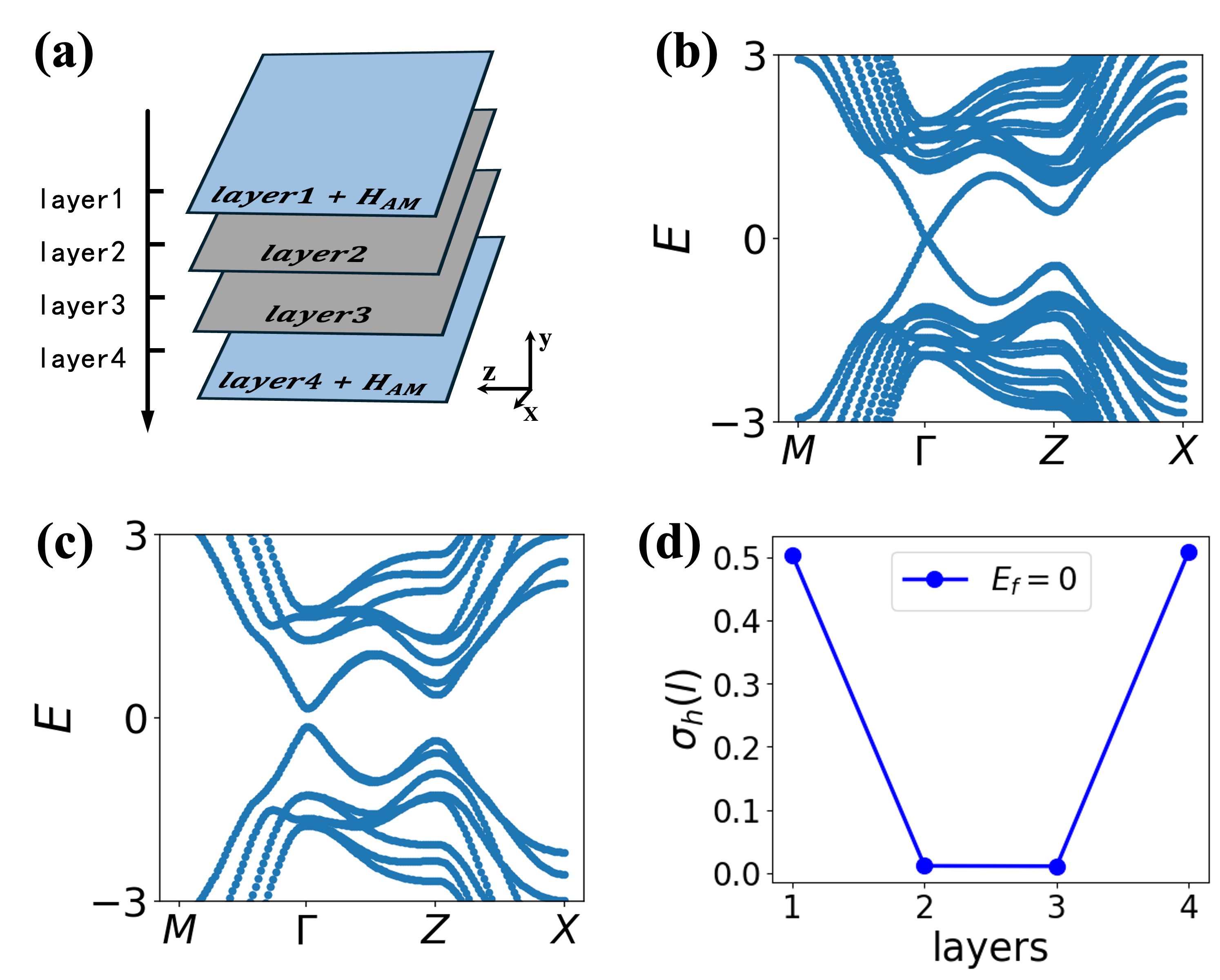}
        \caption{(a) Schematic of a four-layer WTI slab with altermagnetic order on the top and bottom layers. (b) $(010)$ surface band spectrum of an eight-layer slab showing a gapless Dirac cone at $\Gamma$ point and a gapped cone at $Z$ point. (c) Corresponding spectrum for the four-layer slab, where surface hybridization opens a gap at $\Gamma$. (d) Layer-resolved Hall conductance $\sigma_h(l)$ at $E_F = 0$: the outer layers exhibit nearly half-quantized values, while the inner layers have almost zero Hall conductance, yielding a total Hall conductance of $\sigma_H =1 (e^2/h)$. All results were obtained for an altermagnetic strength $J=-0.3$.
    }
        \label{fig3}
\end{figure}

\textit{Four-layer slab model.—}\label{sec3}
We construct a four-layer slab model of a 3D WTI with altermagnetic order on the top and bottom layers, as shown in Fig.\hyperref[fig3]{3(a)}. Unlike the thin-film model, this slab Hamiltonian accounts for finite thickness, enabling access to both bulk and surface states. The tight-binding Hamiltonian for a 3D TI on a simple cubic lattice is taken as\cite{PhysRevB.109.045418}:
 $H_{TI}
 = \Bigl[m - 2\sum_{\alpha=x,y,z} \tilde{B}_\alpha \bigl(1-\cos k_\alpha\bigr)\Bigr]
 s_z \otimes \sigma_0
 +\sum_{\alpha=x,y,z} A\sin k_\alpha s_x \otimes \sigma_\alpha $,
where $m, \tilde{B}_\alpha$ are the model parameters, $s_{x/y/z}$ are the Pauli matrices in the orbital space and $k_\alpha$ is the momentum with $\alpha = x, y, z$. Unless otherwise specified, we choose $A=1, \tilde{B}_x=\tilde{B}_y=1, \tilde{B}_z=0.2$, and $m=2$, which place the system in the WTI phase with $\mathbb{Z}_4$ index $(0,001)$. Based on the Hamiltonian $H_{TI}$ introduced above, we design a four-layer slab stacked along the $y$ ($010$) direction. The corresponding slab Hamiltonian is written as
\begin{equation}
H_{slab}=
\begin{bmatrix}
H_{lay_1}+H_{AM} & D          & 0          & 0 \\
D^{\dagger}      & H_{lay_2} & D    & 0 \\
0                & D^{\dagger}      & H_{lay_3} & D \\
0                & 0                & D^{\dagger} &H_{lay_4}+H_{AM}
\end{bmatrix},
\end{equation}
and the Hamiltonian for each layer is given by
$ H_{lay_l}=
A\sin k_x s_x \otimes \sigma_x
+A\sin k_z s_x \otimes \sigma_z
+\Bigl[m - 2\tilde{B}_y + 2\tilde{B}_x(\cos k_x -1)+2\tilde{B}_z(\cos k_z -1)\Bigr]s_z \otimes \sigma_0 $, where $l$ is the layer index. The interlayer hopping is given by $D=\frac{A}{2i}s_x \otimes \sigma_y+\tilde{B}_ys_z \otimes \sigma_0$. $H_{AM}=J(\cos k_x-\cos k_z)s_0 \otimes \sigma_y $
describes the altermagnetic order applied to the top and bottom surfaces.


In Figs.\hyperref[fig3]{3(b)-3(c)}, we present the band spectrum on the ($010$) surface obtained from the slab model. To highlight the role of surface-surface hybridization, Fig.\hyperref[fig3]{3(b)} shows an eight-layer slab whose top and bottom layers carry altermagnetic order; apart from the increased thickness, all parameters are identical to those of the four-layer slab model. In this case the hybridization between the two surfaces is weak, thus a gapless Dirac cone persists at the $\Gamma$ point, while the Dirac cone at the $Z$ point is gapped by the altermagnetic term. The corresponding ($010$) surface spectrum of the four-layer slab is displayed in Fig.\hyperref[fig3]{3(c)}. Here, the stronger hybridization between the top and bottom surfaces successfully opens a gap at the $\Gamma$ point Dirac cone.

Fig.\hyperref[fig3]{3(d)} shows the layer-resolved Hall conductance (see Supplemental Material~\cite{SupplementalMaterial} (see also references~\cite{PhysRevB.98.245117} therein), Sec.~S4), which is largely insensitive to the choice of Fermi energy, as long as $E_F$ remains inside the bulk and surface gaps. 
Due to the altermagnetic order on the top and bottom surfaces breaking time-reversal symmetry, the outermost layers acquire a finite Hall response, forming an almost half-quantized plateau $(\sigma_h(l)= \frac{1}{2} (e^2/h))$. In contrast, the Hall conductance of the two inner layers is nearly zero. The total Hall conductance therefore forms a quantized plateau, $\sigma_h^{\text{total}} = \sum_l \sigma_h(l) = 1 (e^2/h)$, consistent with the thin-film model results and indicating an overall Chern number $\mathcal{C} = 1$, with the top and bottom surfaces each contributing approximately $1/2$.
The representative value $J=-0.3$ lies well inside the $\mathcal{C}=1$ region of the slab phase diagram; its reasonable energy scale is estimated in Supplemental Material~\cite{SupplementalMaterial} (see also references~\cite{Lee2024MnTeKramers,Li2024MnBi2Te4Review} therein), Secs.~S5 and S6.
We further computed the spin-current counterpart and found no finite spin Hall conductance; see Supplemental Material~\cite{SupplementalMaterial}, Sec.~S7.

\begin{figure}[htbp]
    \includegraphics[width=1\linewidth]{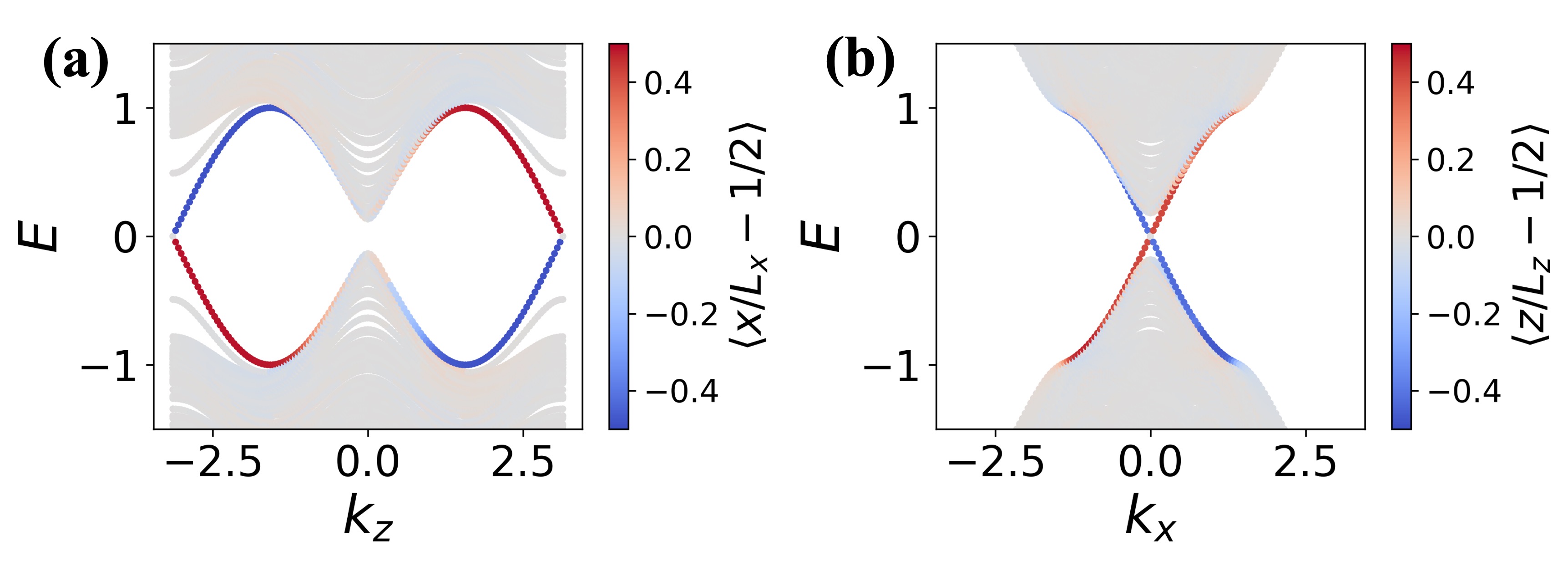}
        \caption{Quasi-1D spectra of the four-layer altermagnetic WTI slab. (a) Spectrum for a slab of size $L_x=25$ and thickness $L_y=4$; the color scale indicates localization on the $(100)$ and $(\bar{1}00)$ side surfaces. A nontrivial gap near $k_z \simeq \pi$ hosts chiral edge modes, while the gap at $k_z = 0$ is trivial. (b) Spectrum as a function of $k_x$ with $L_z=25, L_y=4$; the color encodes localization on the $(001)$ and $(00\bar{1})$ surfaces, where a chiral edge mode appears at $k_x = 0$. All results were obtained for $J = -0.5$.
    }
        \label{fig4}
\end{figure}

We now consider a geometry that preserves translational symmetry only along one direction, which enables us to directly probe the behavior of the edge states. In Fig.\hyperref[fig4]{4(a)}, we show the quasi-1D band spectrum with $k_z$ as a good quantum number (see Supplemental Material~\cite{SupplementalMaterial}, Sec.~S8). Around $k_z \simeq \pi$, a band gap opens and a pair of chiral edge modes appear, localized on the $(100)$ and $(\bar{1}00)$ side surfaces, as indicated by the color scale. This gap is generated by the altermagnetic order on the top and bottom surfaces. At $k_z = 0$, by contrast, the spectrum displays a trivial gap without edge states, arising from the hybridization between the two surfaces.
Fig.\hyperref[fig4]{4(b)} presents the band spectrum with $k_x$ as a good quantum number. At $k_x = 0$, a nontrivial gap opened by the altermagnetic order hosts chiral edge modes localized on the ($001)$ and $(00\bar{1})$ surfaces. Taken together with the band-structure analysis and the layer-resolved Hall conductance, these edge modes demonstrate that the 3D WTI system realizes QAH effect driven by the combined effect of altermagnetic order and surface hybridization.

\begin{figure}[htbp]
    \includegraphics[width=1\linewidth]{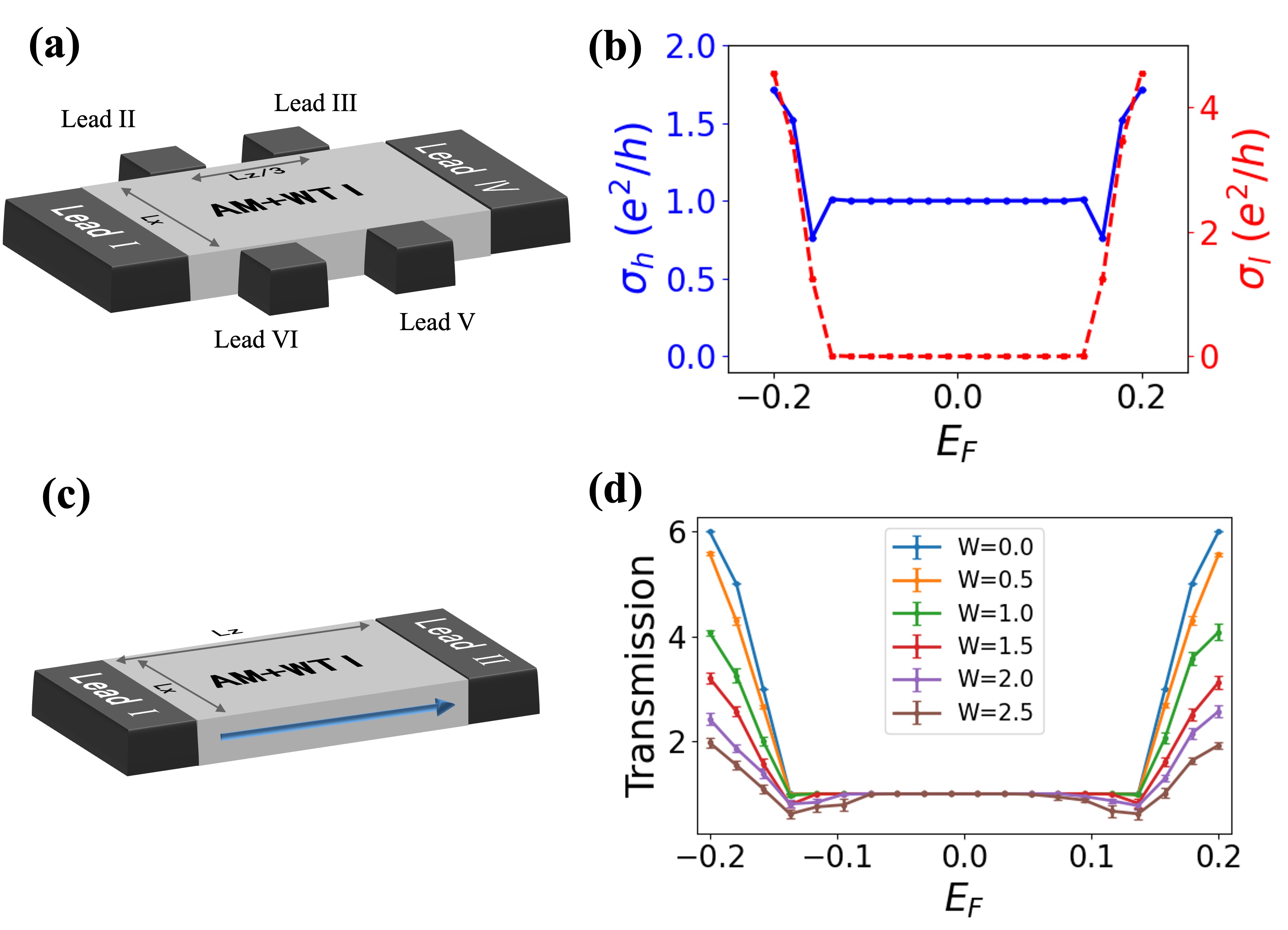}
        \caption{(a) Schematic of the six-terminal Hall-bar geometry based on the altermagnetic WTI slab, where leads I and IV serve as current terminals and leads II, III, V, and VI are voltage probes. (b) Hall conductance $\sigma_h$(blue) and longitudinal conductance $\sigma_l$ (red) as functions of the Fermi energy $E_F$ with $J = -0.5$. (c) Two-terminal bar geometry used to study the effect of Anderson-type potential disorder. (d) Total transmission for different disorder strengths $W$ with $J = -0.5$
    }
        \label{fig5}
\end{figure}

\textit{Transport Properties and Disorder Effects. —}\label{sec4}
To connect the above topological characterization with experimentally accessible observables, we study transport signatures of the QAH phase in an altermagnetic WTI. In contrast to magnetized
strong TIs, transport properties and disorder robustness in WTIs remain largely unexplored. We therefore perform two-terminal and six-terminal calculations and systematically assess disorder effects, providing guidance for experimental detection of the QAH state with no macroscopic magnetization.
Experimentally, the Hall conductance is measured in a six-terminal Hall bar. Accordingly, we simulate transport using the four-layer altermagnetic WTI slab introduced above with $L_x=L_z=50$ and $L_y=4$. As shown in Fig.~\hyperref[fig5]{5(a)}, leads I and IV serve as current leads, while II, III, V, and VI act as voltage probes placed on the side surfaces; the two probe pairs are separated by $L_z/3$ (center-to-center) along the transport direction.

We employ the Landauer-Büttiker formula in the six-terminal Hall-bar geometry. At zero temperature, the current flowing into terminal $n$ is given by \cite{lambert_multi-probe_1993, PhysRevB.109.115305, PhysRevB.84.075416}:
\begin{equation}
    I_n = \frac{e^2}{h}
\sum_{m\neq n}
\bigl[T_{mn}V_n - T_{nm}V_m\bigr],
\end{equation}
where $V_m$ is the voltage at lead $m$, and $T_{mn}(E_F)$ denotes the transmission coefficient at the Fermi energy $E_F$ from lead $n$ to lead $m$. The transmission coefficients are obtained from the Green’s functions of the central scattering region as $T_{mn}(E) =
\mathrm{Tr}\left[\Gamma_m(E) G^R(E)\Gamma_n(E) G^A(E)\right]$, where $\Gamma_n(E) = i\bigl[\Sigma_n^R(E)-\Sigma_n^A(E)\bigr]$ is the line-width function of lead $n$. The retarded and advanced Green’s functions are given by $G^{R(A)}(E)=\bigl[
(E \pm i\eta)\mathbb{I}-H-\sum_n \Sigma_n^{R(A)}(E)\bigr]^{-1}$. In our setup, the left and right leads, which serve as current source and drain, are modeled as semi-infinite perfect electrodes. Their self-energies $\Sigma_\alpha^{R}(E)=\left[\Sigma_\alpha^{A}(E)\right]^{\dagger}$
are computed from the corresponding surface Green’s functions $\Sigma_{\alpha}^R(E) = \tau_\alpha^\dagger g_\alpha^R(E)\tau_\alpha$ $(\alpha = L, R)$, where $g_\alpha^R(E) $ is the surface Green’s function of the semi-infinite lead and $\tau_\alpha$ is the coupling between the lead and the central scattering region.
The remaining terminals are treated as ideal voltage probes within the wide-band limit, and their self-energies are given by $\Sigma_p^R = -\frac{i}{2}\Gamma_p \mathbb{I}$ with the unit matrix $\mathbb{I}$  \cite{PhysRevB.77.115346}(set $\Gamma_p=0.6$ in the letter).


Figure~\hyperref[fig5]{5(b)} shows the Hall and longitudinal conductances (see Supplemental Material~\cite{SupplementalMaterial}, Sec.~S9) versus Fermi energy $E_F$. When $E_F$ lies inside the 
bulk gap, the Hall conductance $\sigma_h$ forms a quantized plateau while the longitudinal conductance $\sigma_l$ is essentially zero. Outside the gap, the plateau collapses: $\sigma_h$ deviates from quantization and $\sigma_l$ rises rapidly. This reflects QAH transport dominated by a single chiral edge channel on the side surfaces (total Chern number $\mathcal{C}=1$), where backscattering is forbidden in the gap. Once $E_F$ enters the bulk bands, additional extended states enable scattering between opposite edges, leading to nonquantized $\sigma_h$ and finite $\sigma_l$.

To probe disorder robustness, we consider Anderson-type potential disorder in a two-terminal geometry to avoid multi-terminal complications. We add a scalar term $V(\mathbf r)s_0\otimes\sigma_0$ in the center region (the gray region in Fig.~\hyperref[fig5]{5(c)}), with $V(\mathbf r)$ independently and uniformly distributed in $[-W/2,W/2]$ (disorder strength $W$). The device in Fig.~\hyperref[fig5]{5(c)} uses the same geometry and parameters as the six-terminal Hall bar, with blue lines indicating the chiral edge states. Figure~\hyperref[fig5]{5(d)} plots the total transmission $T(E_F)$ for several $W$: $T(E_F)\approx 1$ throughout the gap, and the plateau remains essentially intact as $W$ increases, aside from moderate fluctuations near the gap edges. In contrast, outside the gap the transmission is rapidly suppressed with increasing $W$. These results demonstrate that the chiral edge channel in the altermagnetic 3D WTI is highly robust against 
disorder. Here $W$ is measured in the tight-binding energy unit. The range $0\leq W\leq 2.5$, comparable to the hopping/Wilson scales for $A=\tilde{B}_x=\tilde{B}_y=1$, should therefore be viewed as a numerical stress test of the chiral edge channel rather than a material-specific disorder calibration~\cite{PhysRevB.111.045409,cpp8-bgz5}.

Overall, the coexistence of a quantized Hall plateau, vanishing longitudinal conductance, and near-unit transmission, together with their stability over a wide disorder range, provides a clear and experimentally accessible transport signature of a topologically protected QAH phase in the no macroscopic magnetization system.

\textit{Conclusions. —} We have shown that a QAH phase can arise in a WTI through altermagnetic exchange, without net magnetization or magnetic doping. Inter-surface hybridization opens a trivial gap at $\Gamma$, while the altermagnetic exchange gaps the $Z$ cone and generates a $\mathcal{C}=1$ state with a single chiral edge channel and a quantized Hall plateau. Transport calculations reveal strongly suppressed longitudinal response and robustness against disorder.
This mechanism removes the need for a uniform ferromagnetic Zeeman field and avoids magnetic domains, stray fields, and random magnetic dopants.

From an experimental perspective, a natural realization is a WTI/altermagnet heterostructure. Layered bismuth-halide topological materials can supply WTI phases or related boundary states~\cite{Noguchi2024Bi4Br2I2,Zhao2023Bi4Br4,Tang2014Bi2TeI}, while MnTe, CrSb, RuO\(_2\), and Mn\(_5\)Si\(_3\) provide candidate altermagnetic proximity sources~\cite{Krempasky2024MnTeAltermagnetism,Yang2025CrSbAltermagneticSplitting,feng_anomalous_2022,PhysRevB.109.224430}. In this setting, the altermagnetic proximity effect can imprint a momentum-dependent spin splitting onto adjacent nonmagnetic WTI surface layers, effectively producing the exchange term used in our model~\cite{kqy8-myz1}. Such heterostructures therefore offer a plausible route to magnetization-free QAH transport.
%

\textit{Acknowledgments—}This work was financially supported by National Natural Science Foundation of China (Grant Nos. 12474057, 12104028, 124B200091 and 12374034), 
the National Key Research and Development Program of China (Grant No. 2024YFA1409002),
the Quantum Science and Technology-National Science and Technology Major Project (Grant No. 2021ZD0302403),
and the Fundamental Research Funds for the Central Universities.
We also acknowledge the High-performance Computing Platform of Peking University for providing computational resources.\\

\appendix

\bibliography{main}

\end{document}